\documentclass[english]{article}
\usepackage[T1]{fontenc}
\usepackage[latin1]{inputenc}

\makeatletter
 \newenvironment{lyxcode}
   {\begin{list}{}{
     \setlength{\rightmargin}{\leftmargin}
     \setlength{\listparindent}{0pt}
     \raggedright
     \setlength{\itemsep}{0pt}
     \setlength{\parsep}{0pt}
     \normalfont\ttfamily}%
    \item[]}
   {\end{list}}

\usepackage{babel}
\makeatother
\begin{document}

\title{Log Analysis Case Study Using LoGS }

\author{Dmitry Mogilevsky\\
National Center for Supercomputer Applications\\
dmogilev@ncsa.uiuc.edu}

\maketitle
\begin{abstract}
A very useful technique a network administrator can use to identify
problematic network behavior is careful analysis of logs of incoming
and outgoing network flows.  The challenge one faces when attempting
to undertake this course of action, though, is that large networks
tend to generate an extremely large quantity of network traffic in
a very short period of time, resulting in very large traffic logs
which must be analyzed post-generation with an eye for contextual
information which may reveal symptoms of problematic traffic%
\footnote{Such as non-obvious port scans and traffic indicative of worm activity%
}.  A better technique is to perform real-time log analysis using a
real-time context-generating tool such as LoGS.
\end{abstract}

\section{Introduction}

One of the simplest and most comprehensive intrusion detection methods
available to a network administrator is analysis of firewall or network
manager (such as Cisco's NetFlow) logs\cite{Ranum04}. Even logs containing
very basic information, such as destination and source IP address-port
pairs can under proper scrutiny reveal important information about
attempted attacks on the system and internal problems. Firewall or
another central point for network traffic is an ideal spot for collecting
this information, since 

\begin{itemize}
\item All network traffic passes through it.
\item Most firewalls generate excellent activity logs\cite{Cid04}.
\end{itemize}
The usefulness of this technique is diminished by the fact that on
large networks, any interesting messages will be interspersed among
a large number of non-malicious traffic. Combined with the fact that
large networks send and receive on millions of connections per second,
doing intrusion detection becomes a problem of finding a needle in
a hay stack. For example, consider the following fictional log entries:

\begin{lyxcode}
Nov~5~14:03:33,{*}.{*}.{*}.10:3434,1.2.3.5:12346~

Nov~5~14:15:13,{*}.{*}.{*}.10:3434,1.2.3.6:12346~

Nov~5~14:28:32,{*}.{*}.{*}.10:3434,1.2.3.7:12346~

Nov~5~14:40:11,{*}.{*}.{*}.10:3434,1.2.3.8:12346
\end{lyxcode}
This example shows a host on address {*}.{*}.{*}.10 scanning our network
for open NetBus ports\cite{ISS98}. However, since this is a very
slow scan (individual connections are approximately 12 minutes apart),
there may be several million non-informative log entries separating
these interesting entries. Depending on the way the log is audited
by the administrator, it is possible for this trend to be missed.
Instead of analyzing large network logs in search of tell-tale signs
of attempted (or successful) intrusion, the network administrator
may opt for using a real-time log analysis tool to analyze log messages
as they are generated and detect problems based on context created
by previously seen messages.

\section{The Problem}

For this case study, anonymized NetFlow log generated by connections
to and from the network of a large university was examined. The log
is over 1 GB in size. Chronologically, it spans a little over one
hour. It contains 13.6 million individual entries. Currently, these
logs are collected, but not analyzed, despite containing a wealth
of information. It is possible to write a rudimentary script which
will process such logs post-creation and search for interesting information.
In adopting this strategy a number of questions must be considered.

\begin{enumerate}
\item How do I subdivide the log into separate files?
\item How do I collect and store contextual information that will help me
detect port scans?
\item Can I handle potential context overflow from one log to the next?
\item Do I process each log file after its created, or do I collect some
number of log files and process them as a batch?
\end{enumerate}
Appendix 1 shows a sample script to process such logs. While it is
fairly concise, it does a poor job of addressing contextual issues.
Specifically, it only stores state for the last connection, which
makes it impossible for this script to detect any port scans where
the connections are not immediately subsequent. While it is possible
to increase the depth of stored context, doing so would necessitate
creating a store-and-search infrastructure for context, which, when
done improperly can greatly reduce the performance of the script.
Additionally, this script does not address the problem of context
overflowing from one log file to another. These questions can be bypassed
altogether, on the other hand, by real-time scanning with LoGS.

\section{LoGS}

LoGS, currently under development by James Prewett at the Center for
High Performance Computing at University of New Mexico is a highly
customizable and extensible real-time log analysis engine written
in Common Lisp. It is quite efficient, able to process as many as
72,000 messages per second \cite{Pre05}. Because it uses Common Lisp
in its rule definition, it can be programmed to create and store states
and run complex scripts whenever a message is matched \cite{Pre04}. 

The use of Common Lisp makes LoGS unique from other log analysis tools.
Common Lisp was chosen for its flexibility, availability and ease
of use. Lisp also has a fast regular expression engine\cite{Wei03},
which allows LoGS to achieve its high message processing speed. Also,
since Common Lisp is used to configure the rulesets as well, LoGS
is user-extendable.

LoGS consists of five components - Messages, Rules, Rulesets, Actions
and Contexts. Rules associate Messages (input from the analyzed log)
with Actions (Lisp or external scripts). Rulesets extend Rules by
grouping them together into related sets. Contexts collect related
messages together to be processed as a group. Because LoGS actions
can create new rules, as well as update existing rules, LoGS is run-time
configurable. \cite{Pre04}

\section{Analyzing Firewall logs with LoGS}

The context-oriented design of LoGS fits perfectly for the goal of
real-time firewall log analysis. Firewall logs differ from, for example,
system logs in that you cannot implement artificial ignorance\cite{Ranum04},
as every message may potentially contain interesting information when
taken in a context of other messages. Every incoming connection must
be examined against existing contents, and either update them if necessary
or create new ones.

\subsection{Rules}

To effectively detect suspicious behavior in connection logs, LoGS
must be configured to match every incoming message, isolating three
crucial pieces of information - local address, remote address, and
local port. The local port is then checked against a list of vulnerable
ports to identify a possible vulnerability scan. Next, the remote
address, local address and local port are checked against the existing
contexts for matches on remote address and local address (indicating
a potential vertical port scan) or remote address and local port (indicating
a potential horizontal port scan)\cite{Lis}. If an existing context
is found, the message is added to that context. Otherwise a new context
is created for the new message.

\subsection{Contexts}

Each log entry is entered into a context, either a pre-existing one,
or a new one, depending on whether a similar message has been previously
seen. Each context that exists has a timeout, which is incremented
each time a new message is added to the context. When context timeout
occurs, a context action is triggered and the context is removed from
the system. By increasing the size of the timeout value, the administrator
can detect slower port scans (at the price of performance).

\subsection{Actions}

When a context times out and certain conditions are met (for example,
number of accumulated messages in the context. It should be at least
two or more), an action will be triggered. Since LoGS allows to define
actions with arbitrary Lisp programs and even external scripts, its
possible for contexts to trigger very complex series of actions. At
the very least, all the messages in the context should be written
to a separate port scans file. A real-time alert should also be displayed
to the console whenever there is an open context which has accumulated
two or more messages.

\subsection{Defining real-time log analyzer in LoGS}

Using the mentioned elements of LoGS it is fairly easy to construct
a small ruleset to do detailed real-time log analysis. The following
code is a sample of the LoGS code used to accomplish such a task,
written in Common Lisp. 

\begin{lyxcode}

;Firewall~analysis~ruleset~to~spot~vertical

;and~horizontal~port~scan

(make-instance

~~'rule

~~:match

~~(lambda~(message)

~~~~(multiple-value-bind~(matches~sub-matches)

~~~~~~~~(cl-ppcre::scan-to-strings

~~~~~~~~~(\char`\"{}({[}0-9{]}+),({[}0-9{]}+),({[}0-9{]}+:{[}0-9{]}+:{[}0-9{]}+),

~~(TCP|UDP|ICMP),({[}0-9{]}+.{[}0-9{]}+.{[}0-9{]}+.{[}0-9{]}+):({[}0-9{]}+|-{}-),

~~({[}0-9{]}+.{[}0-9{]}+.{[}0-9{]}+.{[}0-9{]}+):({[}0-9{]}+|-{}-),

~~{[}0-9{]}+,({[}0-9{]}+)\char`\"{}

~~~~~~~~~~(message~message)))

~~~~~~(when~matches

~~~~~~~~(values

~~~~~~~~~t

~~~~~~~~~`((sub-matches~,sub-matches))))))

~~:actions

~~~(

~~~~~(make-instance~'rule

~~~~~~~~:match

~~~~~~~~(lambda~(message)

~~~~~~~~~~(multiple-value-bind~(matches~sub-matches)

~~~~~~~~~~~~~~~~(cl-ppcre::scan-to-strings

~~~~~~~~~~~~~~~~~~~~~~~~\char`\"{}({[}0-9{]}+),({[}0-9{]}+),

~~~~~~~~~~~~~~~~~~~~~~~~~({[}0-9{]}+:{[}0-9{]}+:{[}0-9{]}+),

~~~~~~~~~~~~~~~~~~~~~~~~~(TCP|UDP|ICMP),

~~~~~~~~~~~~~~~~~~~~~~~~~(aref~sub-matches~4):({[}0-9{]}+|-{}-),

~~~~~~~~~~~~~~~~~~~~~~~~~(aref~sub-matches~6):({[}0-9{]}+|-{}-),

~~~~~~~~~~~~~~~~~~~~~~~~~{[}0-9{]}+,({[}0-9{]}+)\char`\"{}

~~~~~~~~~~~~~~~~~~~~~~~~(message~message))

~~~~~~~~~~~~~~~~(when~matches

~~~~~~~~~~~~~~~~~~(values

~~~~~~~~~~~~~~~~~~~~T

~~~~~~~~~~~~~~~~~~~~`((sourceip~,(aref~sub-matches~4))

~~~~~~~~~~~~~~~~~~~~~~(sourceport,(aref~sub-matches~5))

~~~~~~~~~~~~~~~~~~~~~~(destip,~(aref~sub-matches~6))

~~~~~~~~~~~~~~~~~~~~~~(destport,~(aref~sub-matches~7)

~~~~~~~~~~~~~~~~~~~~~~(time,~(aref~sub-matches~2))

~~~~~~~~~~~~~~~~~~~)))))

~~~~~~~~:actions

~~~~~~~~(list

~~~~~~~~~~(lambda~(message)

~~~~~~~~~~~~~~~~(declare~(ignore~message))

~~~~~~~~~~~~~~~~(ensure-context

~~~~~~~~~~~~~~~~~~:name~(format~()~\char`\"{}vertical~scan~from~\textasciitilde{}A\char`\"{}~sourceip)

~~~~~~~~~~~~~~~~~~:timeout~(+~get\_universal\_time~timeout\_value)

~~~~~~~~~~~~~~~~~~:actions

~~~~~~~~~~~~~~~~~~(list

~~~~~~~~~~~~~~~~~~(lambda~(message)

~~~~~~~~~~~~~~~~(add-to-context

~~~~~~~~~~~~~~~~~~(format~()~\char`\"{}Vertical~scan:~\textasciitilde{}A:\textasciitilde{}A~to~\textasciitilde{}A:\textasciitilde{}A~at~\textasciitilde{}A\char`\"{}~

~~~~~~~~~~~~~~~~~~~~sourceip~sourceport~destip~destport~time)

~~~~~~~~~~~~~~~~~~message))))

~~~)

(make-instance~'rule

~~~~~~~~:match

~~~~~~~~(lambda~(message)

~~~~~~~~~~(multiple-value-bind~(matches~sub-matches)

~~~~~~~~~~~~~~~~(cl-ppcre::scan-to-strings

~~~~~~~~~~~~~~~~~~~~~~~~\char`\"{}({[}0-9{]}+),({[}0-9{]}+),

~~~~~~~~~~~~~~~~~~~~~~~~~({[}0-9{]}+:{[}0-9{]}+:{[}0-9{]}+),

~~~~~~~~~~~~~~~~~~~~~~~~~(TCP|UDP|ICMP),

~~~~~~~~~~~~~~~~~~~~~~~~~(aref~sub-matches~4):({[}0-9{]}+|-{}-),

~~~~~~~~~~~~~~~~~~~~~~~~~({[}0-9{]}+.{[}0-9{]}+.{[}0-9{]}+.{[}0-9{]}+):

~~~~~~~~~~~~~~~~~~~~~~~~~(aref~sub-matches~7),

~~~~~~~~~~~~~~~~~~~~~~~~~{[}0-9{]}+,({[}0-9{]}+)\char`\"{}

~~~~~~~~~~~~~~~~~~~~~~~~(message~message))

~~~~~~~~~~~~~~~~(when~matches

~~~~~~~~~~~~~~~~~~(values

~~~~~~~~~~~~~~~~~~~~T

~~~~~~~~~~~~~~~~~~~~`((sourceip~,(aref~sub-matches~4))

~~~~~~~~~~~~~~~~~~~~~~(sourceport,(aref~sub-matches~5))

~~~~~~~~~~~~~~~~~~~~~~(destip,~(aref~sub-matches~6))

~~~~~~~~~~~~~~~~~~~~~~(destport,~(aref~sub-matches~7)

~~~~~~~~~~~~~~~~~~~~~~(time,~(aref~sub-matches~2))

~~~~~~~~~~~~~~~~~~~)))))

~~~~~~~~:actions

~~~~~~~~(list

~~~~~~~~~~(lambda~(message)

~~~~~~~~~~~~~~~~(declare~(ignore~message))

~~~~~~~~~~~~~~~~(ensure-context

~~~~~~~~~~~~~~~~~~:name~(format~()~\char`\"{}horizontal~scan~from~\textasciitilde{}A\char`\"{}~sourceip)

~~~~~~~~~~~~~~~~~~:timeout~(+~get\_universal\_time~timeout\_value)

~~~~~~~~~~~~~~~~~~:actions

~~~~~~~~~~~~~~~~~~(list

~~~~~~~~~~~~~~~~~~(lambda~(context)

~~~~~~~~~~~~~~~~~~~~(if~(>=~(get\_universal\_time)~(timeout~context))

~~~~~~~~~~~~~~~~~~~~~~~(report\_context)

~~~~~~~~~~~~~~~~~~)

~~~~~~~~~~~~~~~~~~(lambda~(message)

~~~~~~~~~~~~~~~~(add-to-context

~~~~~~~~~~~~~~~~~~(format~()~\char`\"{}Horizontal~scan:~\textasciitilde{}A:\textasciitilde{}A~to~\textasciitilde{}A:\textasciitilde{}A~at~\textasciitilde{}A\char`\"{}~

~~~~~~~~~~~~~~~~~~~~sourceip~sourceport~destip~destport~time)

~~~~~~~~~~~~~~~~~~message))))

~~~)

~~)

\end{lyxcode}

\section{Conclusions}

Real time firewall log analysis in LoGS offers a flexible and extensible
alternative to batch offline analysis. Its capability for contextual
message parsing is ideally suited for the task of detecting port scans,
as these cannot be detected from any single connection and can only
be inferred by looking at the context of previous connections. LoGS
provides the infrastructure for contextual data collection, and the
capability to trigger arbitrarily sophisticated response. This makes
log analysis a very powerful tool for successful (and potentially
real-time) intrusion detection.

\section{Appendix 1 - Sample Perl code}

\begin{lyxcode}
\#!/usr/bin/perl

print~\char`\"{}Input?~\char`\"{};

\$infile~=~<STDIN>;

open(INPUT,\$infile)||die~\char`\"{}Could~not~open~\$infile\textbackslash{}n\char`\"{};

print~\char`\"{}Outgoing~connections?~\char`\"{};

\$outgoingfile~=~<STDIN>;

print~\char`\"{}Incoming~connections?~\char`\"{};

\$incomingfile~=~<STDIN>;

print~\char`\"{}File~date~(yyyy/mm/dd)?~\char`\"{};

\$date~=~<STDIN>;

print~\char`\"{}Vulnerabilities~file?~\char`\"{};

\$vulfile~=~<STDIN>;

chop(\$date);

\$portscans~=~0;

\$ongoingps~=~0;

open(VULNERABILITIES,~\char`\"{}\$vulfile\char`\"{})

~||die~\char`\"{}Could~not~open~\$vulfile\textbackslash{}n\char`\"{};

open(OUTGOING,\char`\"{}>\$outgoingfile\char`\"{})

~||die~\char`\"{}Could~not~open~\$outgoingfile\textbackslash{}n\char`\"{};

open(INCOMING,\char`\"{}>\$incomingfile\char`\"{})

~||die~\char`\"{}Could~not~open~\$incomingfile\textbackslash{}n\char`\"{};

open(PORTSCANS,\char`\"{}>portscans\_\$infile\char`\"{})

~||die~\char`\"{}Could~not~create~portscans~file\textbackslash{}n\char`\"{};

open(VULSCANS,\char`\"{}>vulscans\_\$infile\char`\"{})

~||die~\char`\"{}Could~not~create~vulnerabilities~scan~file\textbackslash{}n\char`\"{};

while(<VULNERABILITIES>)~\{

~m/\textasciicircum{}({[}a-zA-Z0-9~{]}+),({[}0-9{]}+),(in|out)\$/;

~if~(\$3~==~'in')\{

~~\$invul\{\$2\}~=~\$1~;

~\}

~else\{

~~\$outvul\{\$2\}~=~\$1~;

~\}

\}

\$counter~=~0;

\$sourceip~=~\char`\"{}\char`\"{};

\$sourceport~=~\char`\"{}\char`\"{};

\$destip~=~\char`\"{}\char`\"{};

\$destport~=~\char`\"{}\char`\"{};

\$time~=~\char`\"{}\char`\"{};

while(<INPUT>)~\{

/({[}0-9{]}+),({[}0-9{]}+),({[}0-9{]}+:{[}0-9{]}+:{[}0-9{]}+),

~(TCP|UDP|ICMP),({[}0-9{]}+.{[}0-9{]}+.{[}0-9{]}+.{[}0-9{]}+)

~:({[}0-9{]}+|-{}-),({[}0-9{]}+.{[}0-9{]}+.{[}0-9{]}+.{[}0-9{]}+)

~:({[}0-9{]}+|-{}-),{[}0-9{]}+,({[}0-9{]}+)/;

\$lastsourceip~=~\$sourceip;

\$lastsourceport~=~\$sourceport;

\$lastdestip~=~\$destip;

\$lastdestport~=~\$destport~;

\$lasttime~=~\$time~;

\$time~=~\$3;

\$protocol~=~\$4;

\$sourceip~=~\$5;

\$sourceport~=~\$6;

\$destip~=~\$7;

\$destport~=~\$8;

\$packets~=~\$9~;

if~(\$destport~==~\char`\"{}-{}-\char`\"{})~

~~\{\$destport~=~\char`\"{}0\char`\"{};\}

if~(\$sourceport~==~\char`\"{}-{}-\char`\"{})~

~~\{\$sourceport~=~\char`\"{}0\char`\"{};\}

if~(\$sourceip~=\textasciitilde{}~/10.{[}0-9{]}+.{[}0-9{]}+.{[}0-9{]}+/)\{

~print~OUTGOING~\char`\"{}FWOUT,\$date,\$time~-5:00

~~GMT,\$sourceip:\$sourceport,\$destip:

~~\$destport,\$protocol\textbackslash{}n\char`\"{};

~if~(exists(\$outvul\{\$sourceport\}))\{

~~print~VULSCANS~\char`\"{}Potential~Vulnerability:

~~~\$outvul\{\$sourceport\}.\textbackslash{}n~\$time:~

~~~\$sourceip:\$sourceport~->~\$destip:

~~~\$destport\textbackslash{}n\char`\"{}~;

~\}

\}

else\{

~print~INCOMING~\char`\"{}FWIN,\$date,\$time~-5:00

~~GMT,\$sourceip:\$sourceport,\$destip:

~~\$destport,\$protocol\textbackslash{}n\char`\"{};

~if~(\$sourceip~==~\$lastsourceip~\&\&~

~~~~((\$destip~==~\$lastdestip~\&\&~

~~~~~~\$destport~!=~\$lastdestport)||

~~~~~(\$destport~==~\$lastdestport~\&\&~

~~~~~~\$destip~!=~\$lastdestip)))~\{

~~if~(exists(\$invul\{\$destport\}))\{

~~~~print~VULSCANS~\char`\"{}Potential~Vulnerability:~

~~~~~\$invul\{\$destport\}.\textbackslash{}n~\$time:~

~~~~~\$sourceip:\$sourceport~->~\$destip:

~~~~~\$destport\textbackslash{}n\char`\"{}~;

~~\}

~if~(\$ongoingps~==~0)\{

~\$ongoingps~=~1;

~\$portscans~=~\$portscans~+~1;

~if~(\$destip~==~\$lastdestip)\{

~~print~PORTSCANS~\char`\"{}Potential~vertical~portscan~from~

~~~\$lastsourceip~at~\$lasttime\textbackslash{}n\char`\"{};

~\}

~else~\{

~~print~PORTSCANS~\char`\"{}Potential~horizontal~portscan~from

~~~\$lastsourceip~at~\$lasttime\textbackslash{}n\char`\"{};

~\}

~print~PORTSCANS~\char`\"{}\$lastsourceip:\$lastsourceport~->

~~\$lastdestip:\$lastdestport\textbackslash{}n\char`\"{};

~print~PORTSCANS~\char`\"{}\$sourceip:\$sourceport~->~

~~\$destip:\$destport\textbackslash{}n\char`\"{};

~\}

~else~\{

~print~PORTSCANS~\char`\"{}\$sourceip:\$sourceport~->~

~~\$destip:\$destport\textbackslash{}n\char`\"{};

~\}

\}else~\{

~~\$ongoingps~=~0;

~~\}~

~\}

~\$counter~=~\$counter~+~1~;

~if((\$counter~\%~10000)~==~0)

~\{

~~print~\char`\"{}\$counter\textbackslash{}n\char`\"{};

~\}

\}

print~\char`\"{}\$portscans~Portscans~detected~

~and~written~to~portscans\_\$infile\textbackslash{}n\char`\"{};

close~(INPUT);

close~(OUTGOING);

close~(INCOMING);

close~(VULSCANS);

close~(PORTSCANS);

close~(VULNERABILITIES);
\end{lyxcode}

\end{document}